# A Load Switching Group based Feeder-level Microgrid Energy Management Algorithm for Service Restoration in Power Distribution System

Rongxing Hu, Yiyan Li, Si Zhang, Ashwin Shirsat, Valliappan Muthukaruppan, Wenyuan Tang,
Mesut Baran, David Lubkeman, Ning Lu
Department of Electrical and Computer Engineering
North Carolina State University
Raleigh, NC, USA
{rhu5, yli257, szhang56, ashirsa, vmuthuk2, wtang8, baran, dllubkem, nlu2}@ncsu.edu

*Abstract*—This paper presents a load switching group based energy management system (LSG-EMS) for operating microgrids on a distribution feeder powered by one or multiple grid-forming distributed energy resources. Loads on a distribution feeder are divided into load switching groups that can be remotely switched on and off. The LSG-EMS algorithm, formulated as a mixed-integer linear programming (MILP) problem, has an objective function of maximizing the served loads while minimizing the total number of switching actions. A new set of topology constraints are developed for allowing multiple microgrids to be formed on the feeder and selecting the optimal supply path. Customer comfort is accounted for by maximizing the supply duration in the customer preferred service period and enforcing a minimum service duration. The proposed method is demonstrated on a modified IEEE 33-bus system using actual customer data. Simulation results show that the LSG-EMS successfully coordinates multiple grid-forming sources by selecting an optimal supply topology that maximizes the supply period of both the critical and noncritical loads while minimizing customer service interruptions in the service restoration process.

## I. INTRODUCTION

The number of extreme weather events, such as hurricanes and snowstorms, is climbing in recent years [1]. Those events have caused numerous, widespread long-duration power outages in power distribution systems, making fast service restoration a growing important topic. In the past, distribution feeders without backup generators can only be restored after the main grid power is restored. However, recent technology advancements in grid-forming distributed energy resources (DER), such as hybrid photovoltaic (PV) power plants, where a PV farm has an onsite battery energy storage system (BESS) installed, make it possible to power loads on a distribution feeder by one or a few microgrids.

Many microgrid energy management (EMS) algorithms have been proposed in recent years. In [2], Chen *et al.* proposed an algorithm for recovering the whole distribution system after the weather-related events with abundant DERs. In [3], Xu *et al.* considered the limited capacity of DERs within distribution systems and the unavailability of utility power after major disasters, their objective was to recover critical loads. In [4][5], the authors started to incorporate PV and BESS to aid the system restoration. However, the main energy sources for powering the microgrid are still diesel generators (DGs) or micro-turbines.

There are two main challenges in the feeder-level microgrid EMS design. *First*, if there are multiple grid-forming DERs on a feeder, the microgrid EMS should be able to operate the feeder as one or a few microgrids and select an optimal supply path for each microgrid. However, typical radiality constraints cannot coordinate among multiple grid-forming power sources in a microgrid as pointed out by Wang *et al.* in [6]. To resolve this issue, we propose a new set of topology constraints to incorporate the topology consideration in the optimization problem formulation. *Second*, for a microgrid powered by hybrid PV plants, which have time varying power and energy limits, effective load control becomes a necessity to meet the power and energy supply limits. However, in a typical distribution network, only a limited number of remotely controllable switches are installed for feeder reconfigure. To tackle this problem, we develop a load switching group (LSW) approach for executing microgrid load control.

The contribution of the paper is summarized as follows. *First*, we propose the LSG-EMS as a feeder-level microgrid management algorithm. Formulated as a mixed-integer linear programming (MILP) problem, LSG-EMS has the service restoration based objective function for maximizing the served loads weighted by load priority and minimizing the number of switching actions during the outage period. *Second*, customer comfort and specified restoration requests are accounted for by the minimum service duration constraints and the preferred supply periods. *Third*, a new set of flexible topology constraints are developed to enable optimal supply route selection, allow multiple microgrids to be formed, and allow each microgrid to have more than one power source.

## II. METHODOLGOY

In this section, we will present the LSG-EMS problem formulation.

### A. Assumptions

In this paper, the following assumptions are made: 1) the distributions feeder has only one MW-level hybrid PV plant; 2) the onsite BESS is fully charged at the beginning of the operation; 3) circuit breakers and switches can be controlled by

This research is supported by the U.S. Department of Energy's Office of Energy Efficiency and Renewable Energy (EERE) under the Solar Energy Technologies Office Award Number DE-EE0008770.

the hybrid PV plant remotely; 4) communication remains intact; 5) there is no remotely controllable switches inside an LSG so once an LSG is supplied, all loads inside the LSG are supplied; 6) radial supply topology is strictly maintained at all time so no loop-circuit is allowed to form under any operation conditions; and 7) an LSG containing critical loads has the highest supplying priority.

Note that assumption 7 is proposed because we assume that although all critical loads have onsite backup generators, supplying by an external sources to preserve the onsite generation resources is a preferred operation strategy. For critical loads not following this strategy and operating as islands, we can consider those load nodes as low priority, zero power nodes on the feeder.

*B. Graph Representation of Distribution Feeder Topology*

A graph representation of a simplified distribution feeder is shown in Fig. 1. Each vertex represents either an LSG with DERs (red circles 1 and 6) or an LSG without DER (black circles 2-5). A switch, represented by a rectangular box with red as "on" and white as "off", exists on each edge that connects two vertices. Note that the circuits inside any LSG is radial and cannot remotely be switched on/off by the LSG-EMS.

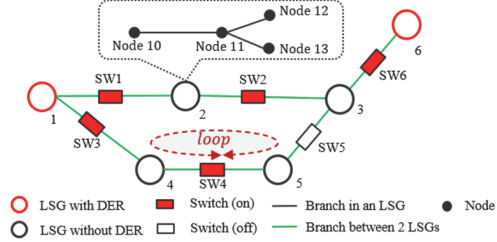

Fig. 1. The schematic connection diagram of LSGs and nodes.

*1) LSG Representation*

For a feeder with a total number of $N_{LSG}$ LSGs, the on/off status of the $m^{th}$ LSG at time $t$ is represented by a binary variable, $U_{m,t}^{LSG}$. Let $\Omega_m^{Node}$ and $\Omega_m^{Branch}$ represent the sets of the load nodes and circuit branches inside the $m^{th}$ LSG, respectively. Note that each entry in a branch set $\Omega_m^{Branch}$ contains a from-to node pair with the first element as the "from" node and the second as the "to" node. The total number of nodes and circuit branches in the $m^{th}$ LSG is $N_m^{Node}$ and $N_m^{Branch}$, respectively.

In Fig. 1, there are 6 LSGs. In LSG2, $\Omega_2^{Node}$ has four load nodes (i.e., nodes 10, 11, 12, and 13) and $\Omega_2^{Branch}$ has three branches represented by three from-to node pairs (10, 11), (11, 12), and (11, 13).

Let binary variables $U_{m,i,t}^{Node}$ and $U_{m,j,t}^{Branch}$ denote the on/off status of the $i^{th}$ load node and the $j^{th}$ branch in the $m^{th}$ LSG at time $t$. Based on assumption 5, we have

$$U_{m,t}^{LSG} = U_{m,i,t}^{Node} = U_{m,j,t}^{Branch}, \quad m = 1 \dots N_{LSG}. \quad (1)$$

*2) Switch Representations*

Let $\Omega_{SW}$ be the set of switches on a distribution feeder. The $n^{th}$ switch controls the on/off of the $n^{th}$ edge that connects two LSGs. Thus, in $\Omega_{SW}$, each switch is represented by a from-to LSG pair, it connects with the first element as the "from" LSG and the second element as the "to" LSG. In Fig. 1, $N_{SW} = 6$, so $\Omega_{SW}$ has 6 from-to LSG pairs (i.e., (1,2), (2,3), (1,4) (4,5), (5,3) and (6,3)) corresponding to the 6 switches.

Represent the on/off status of the $n^{th}$ switch and the two edge connected by the switch at time $t$ by a binary variable, $U_{n,t}^{SW}$ and $U_{n,t}^{Edge}$, respectively, we have

$$U_{n,t}^{SW} = U_{n,t}^{Edge}, \quad n = 1 \dots N_{SW} \quad (2)$$

*3) Topology Constraints*

The spanning tree method (ST) is a widely used radial topology constraints for generating a radial topology from one root [6], which is normally a grid-forming source. A limitation of the existing spanning tree method is that it does not allow multiple generation sources in a radial topology. Therefore, in this paper, we propose a more flexible radial topology constraints that allow multiple generation sources to coexist in the same microgrid.

Our main contribution is to introduce the root status variable for root LSG candidates, $U_{m,t}^r$. If inside an LSG, there are grid forming DERs, the LSG is considered as a root LSG candidate. If the $m^{th}$ LSG is selected as the root for a formed microgrid, $U_{m,t}^r = 1$. All topology constraints are introduced as follows.

Let $m_n^1$ represent the "from" LSG and $m_n^2$ represent the "to" LSG of the $n^{th}$ switch, we have

$$m_n^1 = \Omega_{SW}\{n\}(1); \quad m_n^2 = \Omega_{SW}\{n\}(2). \quad (3)$$

A directional graph is used to represent the edge. Thus, for the $n^{th}$ edge, two binary variables, $\beta_{t,m_n^1,m_n^2}$ and $\beta_{t,m_n^2,m_n^1}$, are used to represent the status and direction of the edge. For example, switch 2 connects LSGs 2 and 3. If $\beta_{t,2,3} = 1$, LSG2 is the parent of LSG3. If $\beta_{t,2,3} = \beta_{t,3,2} = 0$, the edge is disconnected.

Thus, to ensure that the $n^{th}$ switch will only close when either the "from" LSG or the "to" LSG is a parent group:

$$\beta_{t,m_n^1,m_n^2} + \beta_{t,m_n^2,m_n^1} = U_{n,t}^{SW}. \quad (4)$$

Divide the set of LSG groups, $\Omega^{LSG}$, into a set of LSGs with hybrid PV plant, $\Omega^{LSGPV}$, a set of LSGs with DG, $\Omega^{LSGDG}$, and a set of LSGs without DER, $\Omega^{LSGLoad}$, so that

$$\Omega^{LSG} = \Omega^{LSGPV} \cup \Omega^{LSGDG} \cup \Omega^{LSGLoad}. \quad (5)$$

Let $A_m$ devote the set of LSGs which connect to the $m^{th}$ LSG. To guarantee that each LSG without grid-forming resources only have 1 parent LSG or no parent LSG, we have

$$\sum_{x \in A_m} \beta_{t,x,m} = U_{m,t}^{LSG}, \quad m \notin \Omega^{LSGPV} \cup \Omega^{LSGDG}. \quad (6)$$

where $x$ represents the LSG index that connected to LSG $m$.

Conventional ST methods consider all LSGs with generation resources as roots without a parent LSG, so the constraint is represented as

$$\sum_{x \in A_m} \beta_{t,x,m} = 0, \quad m \in \Omega^{LSGPV} \cup \Omega^{LSGDG}. \quad (7\text{-}1)$$

Under this constraint, each generation resource can only form one microgrid and the number of microgrids is fixed, it prevents the EMS algorithm from optimizing microgrid forming options for meeting the supply objectives when there are multiple grid-forming generation resources exist on the feeder.

Therefore, we replace (7-1) by the following set of constraints, which we consider as the main contribution of this paper. First, we introduce the root status variable $U_{m,t}^r$ for root candidates to optimize the root status of the LSGs with generation resources, the number of formed microgrids is determined by the number of the root LSGs, so we have

$$U_{m,t}^{LSG} - U_{m,t}^r \leq \sum_{x \in A_m} \beta_{t,x,m} \leq 0.5(U_{m,t}^{LSG} - U_{m,t}^r + 1),$$
$$m \in \Omega_G^{LSGPV} \cup \Omega_G^{LSGDG}. \quad (7\text{-}2)$$

To guarantee that a root LSG will definitely be served, we have

$$U_{m,t}^r \leq U_{m,t}^{LSG}, \ m \in \Omega^{LSGPV} \cup \Omega^{LSGDG}. \quad (8)$$

To guarantee that a switch is off when either its "from" or "to" LSG is unenergized, we have

$$U_{n,t}^{SW} \leq 0.5(U_{m_n^1,t} + U_{m_n^2,t}). \quad (9)$$

Let $\Omega_O$ represent the set of loops and $\Omega_{SW,c}$ the set of switches in the $c^{th}$ loop. Let $N_{SW}^{O,c}$ be the number of switches in loop $c$, to guarantee that no loop can be formed, the loop constraint can be represented as

$$\sum_{y \in \Omega_{SW,c}} U_{y,t}^{SW} \leq N_{SW}^{O,c} - 1. \quad (10)$$

where $y$ represents the switch index in loop $c$.

The number of the served LSGs is equal to the number of "on" switches and the number of roots

$$\sum_{n \in \Omega_{SW}} U_{n,t} + \sum_{m \in \Omega_G^{LSGPV} \cup \Omega_G^{LSGDG}} U_{m,t}^r = \sum_{m \in \Omega_G} U_{m,t}. \quad (11)$$

### C. LSG-EMS Problem Formulaton

The main decisions for service restoration in a distribution system by microgrid EMSs that coordinate and operate the hybrid PV plant and other grid-forming DGs are to 1) select LSGs to serve, 2) determine the number of microgrid to form, and 3) select the optimal supply path.

Thus, we formulate the objective function of the LSG-EMS algorithm as

$$\min \sum_{t=1}^{T} \sum_{m=1}^{N_{LSG}} \left[ U_{m,t}^{LSG} \sum_{i=1}^{N_m^{Node}} \left( w_{m,i}^{pri} w_{m,i,t}^{pref} P_{m,i,t} \Delta t \right) \right] - k_1 \sum_{t}^{T} \sum_{n}^{N_{SW}} U_{n,t}^{so} \quad (12)$$

where $T$ is the scheduling horizon, $\Delta t$ is the scheduling interval (30 minutes in this paper); $P_{m,i,t}$ is the power consumption of node $i$ in the $m^{th}$ LSG at time $t$; $w_{m,i,t}^{pref}$ is the weighting representing customers' preference for the $t^{th}$ scheduling interval; $w_{m,i}^{pri}$ is the load priority weighting for node $i$ in the $m^{th}$ LSG; $k_1$ represents the weighting of the total number of switching actions; and $U_{n,t}^{so}$ is a binary variable representing the $n^{th}$ switch action at $t$.

Note that although $U_{n,t}^{so}$ can be directly calculated by $U_{n,t}^{so} = |U_{n,t}^{SW} - U_{n,t-1}^{SW}|$, to facilitate the solving of the optimization problem using MILP, we formulate the calculation of $U_{n,t}^{so}$ as

$$U_{n,t}^{SW} - U_{n,t-1}^{SW} \leq U_{n,t}^{so}, \quad (13)$$
$$U_{n,t-1}^{SW} - U_{n,t}^{SW} \leq U_{n,t}^{so}. \quad (14)$$

The first component of (12) is the total amount of load weighted load (by $w_{m,i}^{pri}$ and $w_{m,i,t}^{pref}$) for prioritizing the supply of high priority loads (i.e., with higher $w_{m,i}^{pri}$) and customer preferred supply durations (i.e., supply periods with higher $w_{m,i,t}^{pref}$). The second component in (12) is to minimize the total number of switching operations to avoid unnecessary switching transients, reduce wear-and-tear, and improve system reliability.

#### 1) Customer Comfort Constraints

We propose to include a minimum service duration, $T^{MSD} = K^{MSD} \Delta t$, which will last for $K^{MSD}$ consecutive time steps, to ensure an LSG will be served for a minimum duration. A reasonable minimum service duration for a distribution customer is 2 or 3 hours so that most of the critical needs of a customer can be satisfied. Let $\hat{K}$ be the feasible consecutive time steps if an outage duration is shorter than a given $T^{MSD}$.

$$\hat{K} = \min\{K^{MSD}, T - t + 1\}, \quad (15)$$
$$\sum_{z=0}^{\hat{K}-1} U_{m,t+z}^{LSG} \geq \hat{K}(U_{m,t}^{LSG} - U_{m,t-1}^{LSG}), \ t > 1. \quad (16)$$

#### 2) Voltage Regulation Constraints

If a hybrid PV plant and a few DGs are in the same microgrid, we assume that the hybrid PV plant will regulate the voltage as the main power source. Therefore, the root status of the LSG with the hybrid PV plant equals to its group status. Let $m_{PV}$ be the index of the LSG with hybrid PV plant, $i_{m,PV}$ is the index of the node with hybrid PV plant in LSG $m_{PV}$, we have

$$U_{m_{PV},t}^r V_{rate}^2 \leq V_{i_{m,PV},t}^2 \leq U_{m_{PV},t}^r V_{rate}^2 + M(1 - U_{m_{PV},t}^r), \quad (17)$$
$$U_{m_{PV},t}^r = U_{m_{PV},t}. \quad (18)$$

where $V_{rate}$ is the rated voltage and $M$ is a large number.

#### 3) Other Opreational Constraints

Note that the feeder topology constraints have been presented in Section II B.3). As we do not consider the formulation of conventional operational constraints as our contribution, we provide references for formulating those constraints as follows. For formulating linear DistFlow constraints, please refer to (1)-(6) and (37) in [2]. Status variables of nodes and branches are replaced by the status of LSGs and switches. For formulating the reserve constraints, please refer to (16) in [2]. We set the reserve capacity of the hybrid PV plant and each DG at 15% of the load to be supplied.

For formulating reactive power constraints of inverters of the PV farm, the onsite BESS and the DGs with polygon-based linearization, please refer to (12)-(15) in [2]. Note that we choose the inner octagon approximation of a circle and assume that inverters cannot absorb reactive power. We also oversize the PV inverter at 110% of the rated PV farm capacity.

For DGs associated operational constraints, please refer to (17)-(18) in [2]. For BESS associated operational constraints, please refer to (17)-(20) in [4]. Note that we ignore the ramping rate constraints of the onsite BESS and DGs. We set the minimum and maximum active power of DGs as 25% and 100% of the rated power, respectively. BESS charging/discharging efficiency is 95%. BESS minimum and maximum energy storage is 20% and 100% of its capacity, respectively.

### III. SIMULATION RESULTS

In this paper, to illustrate the setup and quantify the performance of the LSG-EMS algorithm, we build a test system based on the 33-bus system [7]. As shown in Fig. 2, the feeder is divided into 7 LSGs with 11 remotely controlled switches, it has 21 possible supply loops. The hybrid PV plant is located at node 2, and consists of a 2200kW PV farm and a 4000kWh onsite BESS with a rated charging/discharging power of 1000kW. A 400kW DG is installed at node 16.

We assume the feeder microgrid operates under a long outage scenario. So the scheduling period is from 0:00 to 24:00. All switches are assumed to be open at $t = 0$. The two user preferred service time periods are 7:00-9:00 and 18:00-20:00 with $w_{m,i,t}^{pref} = 1.5$. The two critical loads, an industry customer having a peak load at 68.2kW and a medical center having a peak load at 48.5kW, are located at nodes 5 and 10, both with service weighing of $w_{m,i}^{pri} = 2$. Critical Load profiles are 15-minute smart meter data. To populate the load profile on each load node, we randomly select load profiles from smart meter database until the peak of the aggregated load at each load node is more than 110% that of the original 33-bus system [7]. PV profile is selected from a PV farm measurement with 30-minute resolution. The load and PV data are down-sampled to 30-minute, as shown in Fig. 3. $k_1$ is 1.

The LSG-EMS algorithm has been formulated as a MILP problem, it is solved by the CPLEX 12.10 solver called by MATLAB 2019b on a desktop with I9-9900(3.1Ghz) CPU and 64G RAM.

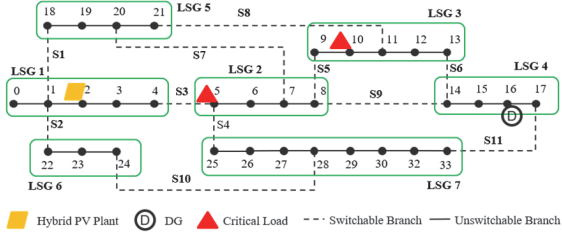

Fig. 2.  The single-line diagram of the modified 33-bus sytem.

### A. Overall Performance

An example of the LSG-EMS scheduling results with $T^{MSD} = 2$ hours is shown in Fig. 3. The runtime is 990 seconds. The onsite BESS charges from 10:00 to 16:00 and there is no PV curtailment in the entire supply duration. When the hybrid PV plant is offline, LSG1 will suffer a power interruption from 20:00 to 22:00 as the hybrid PV plant can no longer satisfy the MSD requirement. The DG in LSG4 generates at the maximum output in the microgrid controlled by the hybrid PV plant. After the hybrid PV plant is offline, the active power output of the DG drops significantly as it can only serve LSG4. The results demonstrate that the LSG-EMS can coordinate multiple grid-forming sources for achieving supply objectives.

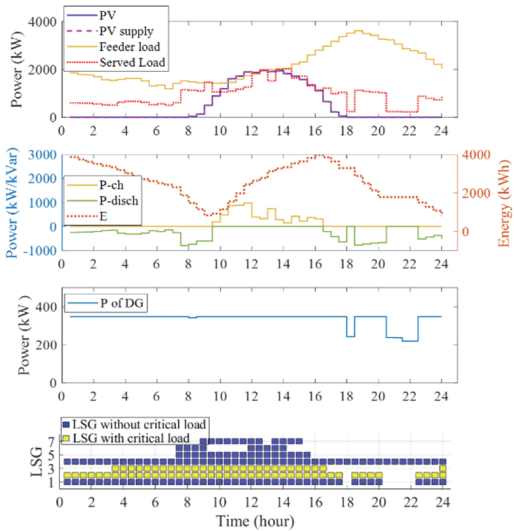

Fig. 3.  PV, BESS, DG and load profiles ($T^{MSD}= 2$ hours).

### B. Voltage Regulation

LSG-EMS operates the hybrid PV plant and DG for providing voltage regulation. As shown in Fig. 4, nodal voltages are maintained within required voltage limits, [0.95 1.05] p.u. In this case, we let the PV plant regulates its voltage at 1 p.u. when online. Note that the voltage of the DG can be higher than 1.0 p.u. when injecting active power in the microgrid.

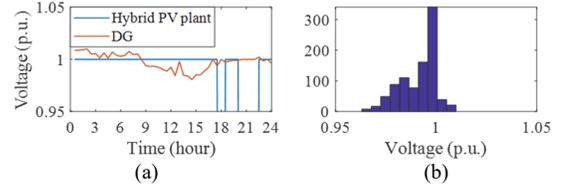

Fig. 4.  (a) Voltages profiles of the hybrid PV plant and DG. (b) Voltage distribution of all served nodes ($T^{MSD}= 2$ hours).

### C. Impact of Minimum Service Duration Selection

In Fig. 5, we show the impact of $T^{MSD}$ on the supply duration of each LSG. As expected, a longer $T^{MSD}$ guarantees that each LSG is served for a longer, consecutively duration. However, serving LSGs during the second customer preferred service period (18:00-20:00) becomes very challenging. This is because between 18:00-20:00, loads in all LSGs are peaking while the PV generation is diminishing. As a result, for $T^{MSD} > 0.5$ hour, only 3 LSGs can be served by a microgrid powered by the onsite BESS and the DG, whereas when $T^{MSD} = 0.5$ hour, LSG5 can also be served briefly.

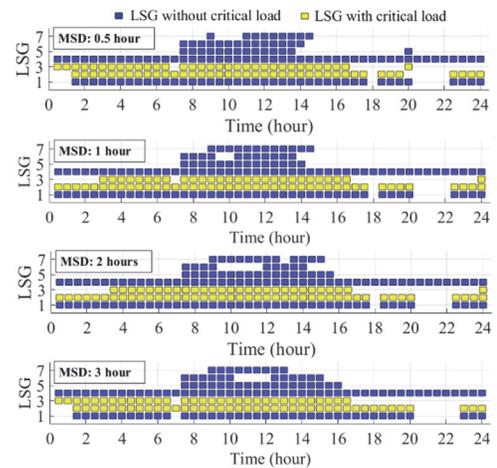

Fig. 5.  Served status of LSGs with different minimum service durations.

TABLE I. RESILIENCE SEVRVICE PERFORMANCE UNDER 4 $T^{MSD}$S

| Minimum Service time $T^{MSD}$ (h) | | **0.5** | **1** | **2** | **3** |
|---|---|---|---|---|---|
| Served Demand (kWh) | | 22,780 | 22,817 | **22,818** | 22,748 |
| Served Time of Critical Load (h) | Node 5 | **20** | 21.5 | 21.5 | 20.5 |
| | Node10 | **16.5** | 14 | 14 | 16 |
| Number of Nodes Served in Preferred Period | 7:00-9:00 | 25 | 25 | 25 | 25 |
| | 18:00-20:00 | **22** | 13 | 13 | 13 |
| Number of Switching Actions | | **28** | 23 | 23 | **22** |

In Table I, we compare the performance of four $T^{MSD}$ settings with values highlighted in green are the best and in red are the worst. The results clearly show that a longer $T^{MSD}$ indeed limit the microgrid supply flexibility (i.e., 18:00 to 20:00) in high demand hours where all loads prefer to be supplied and there are not enough generation resources. However, the longer $T^{MSD}$ can effectively reduce the switching actions. The case when $T^{MSD} = 0.5$ hour suffers more frequent service interruptions than the case with $T^{MSD} = 3$ hours (see Fig. 5). However, the difference among the cases when $T^{MSD}$ varies between 1 and 2 hours is marginal, showing there is a tradeoff only when $T^{MSD}$ exceeds a threshold that is determined by the availability of the power supply and the demand of each LSG in customer preferred supply periods. In subsequent cases, $T^{MSD}$ is set at 2 hours as it exhibits the best overall performance.

*D. Supply Path Selection and Microgrids Forming*

In Fig. 6, all topologies selected by the LSG-EMS during a 24-hour supply period with the PV capacity set at 2200 kW are shown for comparison. From Fig. 6, we can see that when $T^{MSD} = 2$ hours, there are 6 topologies among the 7 topologies selected where the two grid-forming resources can form a single microgrid. This shows that the algorithm successfully maximizes the supply radius by coordinating the two possible root candidates. The results also demonstrate that the LSGs with critical loads have a higher chance to be supplied. However, as all LSGs on a supply path for picking up the critical LSG, will be supplied, LSG2 is more likely to be supplied than LSG3. This is because LSG2 is directly connected to LSG1 while LSG3 is not.

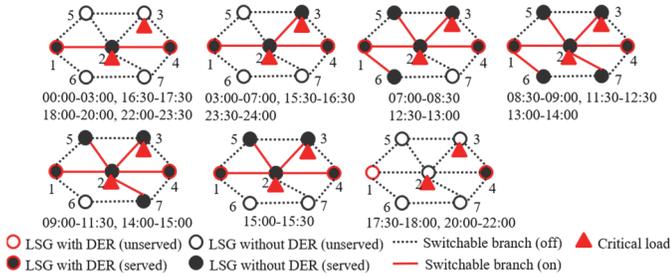

Fig. 6. Circuit topologies during the service restoration ($T^{MSD}$= 2 hours).

In Table II, we present 9 dominant topologies out of 21 topologies selected by the LSG-EMS when we scale the 2200 kW PV up by a capacity factor from 0.5 to 2. The feeder operates under a dominant topology for at least 2.5 hours (i.e. 5 scheduling intervals) in a 24-hour operation period. Fig. 7 shows the supply status of the 7 LSGs under the four PV capacity settings. From Table II, topologies 1 and 3 are selected in all 4 cases while topologies 6, 7, 8 and 9 only occur in certain high PV cases. Topology 4 occurs less often in low PV case but will become dominant in high PV cases while Topology 5 dominates in the low PV cases. Identification of dominant topologies is crucial because once stable dominant topologies are found, LSG-EMS can select supply path and options only from the dominant topologies. This will not only significantly shorten the computing time when there are multiple grid-forming DER exist but also greatly simplify the microgrid protection design by facilitating relay coordination.

## IV. CONCLUSION

In this paper, we present a LSG-based microgrid energy management algorithm for service restoration using multiple grid-forming distributed generators. A set of flexible topology constraints are developed to allow the LSG-EMS to optimize how may microgrids to form and how many DERs should be in each formed microgrid. The LSG-EMS accounts for customer comfort by satisfying demand in customer specified hours and setting a minimum service duration to reduce customer service interruptions. In our follow-up journal paper, we will present more details on dominant topology selection and address the inaccuracy in PV and load forecast on microgrid operation.

TABLE II. DOMINANT TOPOLOGIES UNDER DIFFETEND PV CAPACITIES.

| Dominant Topology | PV Capacity Scale Factor | | | | Dominant Topology | PV Capacity Scale Factor | | | |
|---|---|---|---|---|---|---|---|---|---|
| | 0.5 | 1.0 | 1.5 | 2.0 | | 0.5 | 1.0 | 1.5 | 2.0 |
| 1 | 5 | 5 | 4 | 4 | 6 | - | 7 | - | - |
| 2 | 3 | - | 1 | 1 | 7 | - | - | - | 15 |
| 3 | 21 | 15 | 12 | 10 | 8 | - | 5 | - | - |
| 4 | 1 | - | 11 | 11 | 9 | - | - | 14 | - |
| 5 | 11 | 11 | - | - | | | | | |

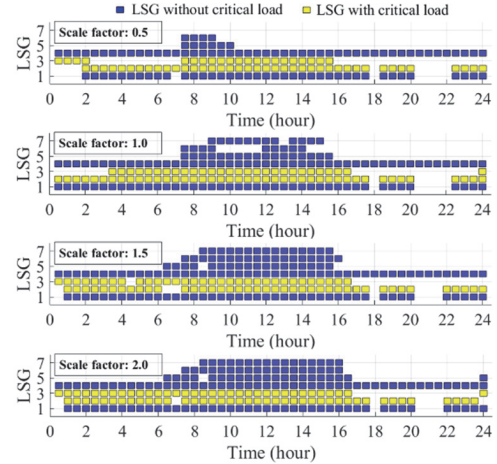

Fig. 7. Served status of LSGs under different PV capacity scale factors.

ACKNOWLEDGMENT

The authors thanks PJ Rhem with ElectriCities, Paul Darden, Steven Hamlett and Daniel Gillen with Wilson Energy for their inputs, suggestions and technical support.